%% file: main.tex
\documentclass[conference]{IEEEtran}
\IEEEoverridecommandlockouts
\usepackage{cite}
\usepackage{amsmath,amssymb,amsfonts}
\usepackage{graphicx}
\usepackage{textcomp}
\usepackage{xcolor}

\usepackage{textcomp}
\usepackage{xcolor}
\usepackage{array}
\usepackage{xspace}
\usepackage{listings}
\usepackage{longtable}
\usepackage{lscape}
\usepackage{threeparttable}

\usepackage[normalem]{ulem}
\usepackage{pifont}
\usepackage{multirow}
\usepackage{algorithm}
\usepackage{algpseudocode}
\usepackage{amsmath, amssymb}
\usepackage{comment}
\usepackage{xspace}
\usepackage{hyperref}
\usepackage{marvosym} 

\newcommand*{\ie}{i.e.,\@\xspace}

\def\BibTeX{{\rm B\kern-.05em{\sc i\kern-.025em b}\kern-.08em
    T\kern-.1667em\lower.7ex\hbox{E}\kern-.125emX}}

\newcommand*{\tool}{\texttt{RAGSum}\@\xspace}
\newcommand*{\JC}{\texttt{JOINTCOM}\@\xspace}
\newcommand*{\CMR}{\texttt{CMR-Sum}\@\xspace}
\newcommand*{\JCSD}{\texttt{JCSD}\@\xspace}
\newcommand*{\PCSD}{\texttt{PCSD}\@\xspace}
\newcommand*{\CCSD}{\texttt{CCSD}\@\xspace}
\newcommand*{\Llama}{\texttt{LLama-3.1-8B}\@\xspace}

\newcommand{\rqone}{\textbf{RQ$_1$}: \emph{How effective is the Retriever component of \tool in retrieving relevant results compared to the baselines?}}

\newcommand{\rqtwo}{\textbf{RQ$_2$}: \emph{How effective is \tool compared to the baselines?}}

\newcommand{\rqthree}{\textbf{RQ$_3$}: \emph{How does each component of \tool contribute to its overall performance?}}

\begin{document}


\title{When Retriever Meets Generator: A Joint Model for Code Comment Generation
}

\author{\IEEEauthorblockN{Tien P. T. Le, Anh M. T. Bui$^{*}$\thanks{*Corresponding author}, Huy N. D. Pham}
\IEEEauthorblockA{
\textit{Hanoi University of Science and Technology}\\
Hanoi, Vietnam \\
tien.lpt207633@sis.hust.edu.vn, anhbtm@soict.hust.edu.vn}
\and
\IEEEauthorblockN{Alessio Bucaioni}
\IEEEauthorblockA{
\textit{Mälardalen University}\\
Västerås, Sweden \\
alessio.bucaioni@mdu.se}
\and
\IEEEauthorblockN{Phuong T. Nguyen}
\IEEEauthorblockA{
\textit{University of L'Aquila}\\
L'Aquila, Italy \\
phuong.nguyen@univaq.it}
}

\maketitle

\begin{abstract}
Automatically generating concise, informative comments for source code can lighten documentation effort and accelerate program comprehension. 
Retrieval-augmented approaches first fetch code snippets with existing comments and then synthesize a new comment, yet retrieval and generation are typically optimized in isolation, allowing irrelevant neighbors to propagate noise downstream. To tackle the issue, we propose a novel approach named \tool with the aim of both effectiveness and efficiency in recommendations. \tool is built on top of fuse retrieval and generation using a single CodeT5 backbone. 
We report preliminary results on a unified retrieval-generation framework built on CodeT5. A contrastive pre-training phase shapes code embeddings for nearest-neighbor search; these weights then seed end-to-end training with a composite loss that \emph{(i)} rewards accurate top-k retrieval; and \emph{(ii)} minimizes comment-generation error. More importantly, a lightweight self-refinement loop is deployed to polish the final output.
We evaluated the framework on three cross-language benchmarks (Java, Python, C), and compared it with three well-established baselines. The results show that our approach substantially outperforms the baselines with respect to BLEU, METEOR, and ROUTE-L. 
These 
findings indicate that tightly coupling retrieval and generation can raise the ceiling for comment automation and motivate forthcoming 
replications and qualitative developer studies.


\end{abstract}


\section{Introduction}
\label{sec:intro}

\input{src/Introduction}

\section{Related Work}
\label{sec:rw}
\input{src/Related_Work}

\section{Proposed Approach}
\label{sec:ProposedApproach}
\input{src/Proposed_Solution}

\section{Evaluation}
\label{sec:settings}
\input{src/Empirical_Evaluation}

\section{Results and Discussion}
\label{sec:result}
\input{src/Result}

\section{Conclusion and Future Work}
\label{sec:conclusion}
\input{src/Conclusion}

\section*{Acknowledgment} 
\small
This work is supported by the Swedish Agency for Innovation Systems through the project ``Secure: Developing Predictable and Secure IoT for Autonomous Systems" (2023-01899), and by the Key Digital Technologies Joint Undertaking through the project ``MATISSE: Model-based engineering of digital twins for early verification and validation of industrial systems" (101140216). This paper has been partially supported by the MOSAICO project that has received funding from the EU Horizon Research and Innovation Action (Grant Agreement No. 101189664). We acknowledge the Italian ``PRIN 2022'' project TRex-SE: \emph{``Trustworthy Recommenders for Software Engineers,''} grant n. 2022LKJWHC. 

\bibliographystyle{IEEEtran}
\bibliography{sample-base}

\end{document}

%% file: src/Introduction.tex
Up-to-date, readable comments accelerate program comprehension, reduce onboarding time and maintenance cost~\cite{xie2021exploiting}. Yet surveys show that 60–70\% of developers routinely encounter missing or obsolete comments~\cite{ahmed2022few}, and such mismatches raise the likelihood of defect-inducing changes by roughly 1.5x. Automating comment generation has therefore become an active line of inquiry at the intersection of software engineering and natural-language processing.

\begin{figure}
    \centering
    \includegraphics[width=0.9\linewidth]{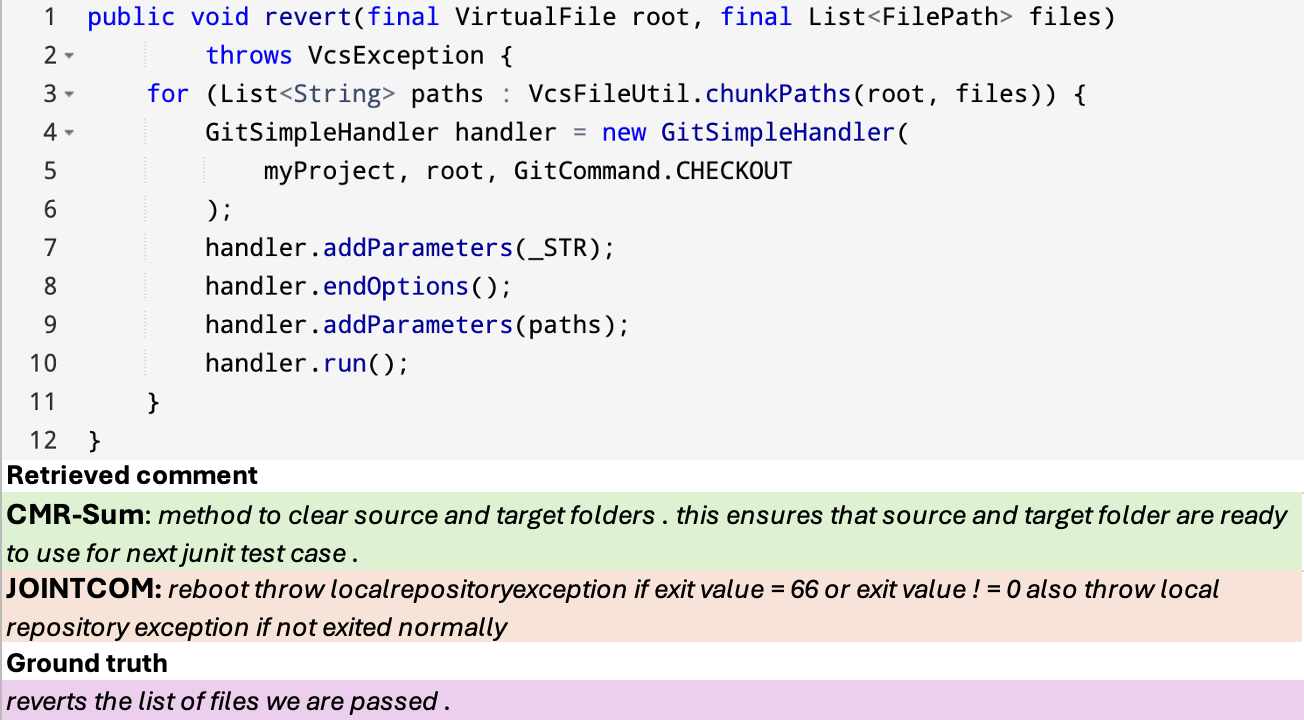}
    \caption{Example of retrieved comments by \CMR and \JC.}
    \label{fig:example}
\end{figure}

Early work tackled the problem with template rules and information-retrieval (IR) heuristics. Template systems extract salient tokens and stitch them into fixed linguistic patterns~\cite{sridhara2010towards,mcburney2015automatic}; IR systems locate code fragments similar to a query and reuse their comments~\cite{wong2013autocomment,wong2015clocom}. Although lightweight, these methods often misalign with the precise semantics of the target snippet.
The advent of neural sequence-to-sequence models re-framed comment generation as a machine-translation task from source code to natural language~\cite{iyer2016summarizing,hu2018deep}. Such models learned richer representations, but even the best variants struggled to bridge the modality gap between programming languages and English, leading to generic or inaccurate summaries~\cite{leclair2019neural}.
To mitigate these weaknesses, recent studies blended IR with neural generation. The model first retrieves code–comment exemplars and then conditions the decoder on that context~\cite{ye2020leveraging,parvez2021retrieval,li2021editsum}.
Although this paradigm has shown promise, existing systems typically train retrieval and generation components separately, which can lead to irrelevant neighbors introducing noise into the generated comments and hinder overall performance.
To address this issue, Li et al.~\cite{li2021editsum} proposed \texttt{EditSum} that refines retrieved comments to better align with the semantics of the input code query. While \texttt{EditSum} captures essential keywords from the input code snippet during comment generation through its self-editing pipeline, the presence of irrelevant retrieved code can still degrade performance.
A joint training approach for simultaneously optimizing the retriever and generator has been employed in \JC~\cite{lu2024improving} and later, \CMR~\cite{li2024cross}, to enhance the retrieval of relevant comments.
These approaches aim to achieve a balance between the two tasks' performance by employing a shared learning framework. 
While \JC treated retrieval and generation as two separate models, sharing weights between them during training; \CMR proposed an extractor that integrates generated and retrieved comments within a unified framework, aiming to align them using an attention mechanism. 
We argue that \emph{though these approaches outperform earlier methods based on separate training paradigms, treating the retriever and generator as distinct tasks may still hinder the overall performance of comment generation}. As a motivating example, in Figure~\ref{fig:example}  we show the results of using \JC and \CMR to generate comments for a given input code query. It is evident that compared to the ground-truth comment and the input code, the results generated 
by both \JC and \CMR exhibit significant semantic inaccuracies.

To bridge such a gap, we \textit{report preliminary results} on a novel model to fuse retrieval and generation within a single CodeT5~\cite{wang2021codet5} backbone, with the aim of both effectiveness and efficiency in the final recommendations. First, an initial contrastive phase shapes the encoder for nearest-neighbour search. Second, a composite objective tunes both encoder and decoder end-to-end, rewarding accurate top-k retrieval and fluent, context-aware comments. Third, a lightweight self-refinement loop further polishes the output. To study \tool, we evaluated it on three cross-language benchmarks, \ie Java, Python, C and compared it with three well-established baselines, \ie \CMR \cite{li2024cross}, \JC \cite{lu2024improving}, and \Llama \cite{grattafiori2024llama}. 
The experimental results showed that \tool gains significant improvements with respect to the baselines.
These early findings indicate that tightly coupling retrieval and generation can raise the ceiling for comment automation and motivate forthcoming industrial replications and qualitative developer studies. 

The main contributions of our work are 
as follows.

\begin{itemize}
    \item \textbf{We developed \tool}, a practical approach to code comment generation using 
    contrastive pre-training phase shapes code embeddings for nearest-neighbor search.
    \item \textbf{We conducted an empirical evaluation} using three real-world datasets to study \tool's performance and compare it with three well-established baselines.
    \item \textbf{We published online a replication package} including the data curated and tool developed through this work 
    to foster future research~\cite{ragsum}.    
\end{itemize}

The paper is organized as follows. In Section~\ref{sec:rw}, we review the related work. Section~\ref{sec:ProposedApproach} explains in detail the proposed approach. The empirical evaluation to study the performance of \tool is presented in Section~\ref{sec:settings}. Afterward, in Section~\ref{sec:result}, we report and analyze the experimental results. Finally, Section~\ref{sec:conclusion} sketches future work, and concludes the paper.


%% file: src/Related_Work.tex
Deep learning can automatically learn pattern features from large-scale datasets, several studies have explored deep learning-based methods for code summarization\cite{wan2018improving}. With the advantage of transformer, sequence-to-sequence (Seq2Seq) architectures bring significant improvements for generating summaries of code. Transformer-based models \cite{feng2020codebert}, \cite{guo2020graphcodebert} have enhanced the semantic understanding of comment generation. Several approach focused on leveraging Abstracted Syntax Tree (AST) as input of encoder-decoder model \cite{shi-etal-2021-cast}.  However, generation models often struggle with issues such as hallucination and limited access to external knowledge, which can hinder the accuracy and completeness of the generated summaries
To address this limitations, Zhang et al. \cite{zhang2020retrieval} introduced \texttt{Rencos}, a retrieval-based neural approach for source code summarization but lack dynamic integration during generation. Another framework for comment generation -- \texttt{DECOM} \cite{mu2022automatic} with the multistage deliberation process which use the keywords from source code and the comment of retrieved sample to enhance the performance. 
However, these approaches treated the retriever and generator as separate components, training them in isolation and thereby limiting their potential synergy. 
Recent studies \cite{lu2024improving, li2024cross} have proposed combining retrievers and generators to leverage their complementary strengths. 
Many research focus on the ability of LLMs in code comment generation. Recent research has concentrated on exploring various prompting techniques to better harness the potential of LLMs in this task \cite{sun2024source} but the summaries produced by LLMs often differ significantly in expression from reference and tend to include more detailed information than those generated by traditional models \cite{sun2023automatic}.



%% file: src/Proposed_Solution.tex
\begin{figure*}[ht]
    \centering
    \includegraphics[width=0.88\textwidth]{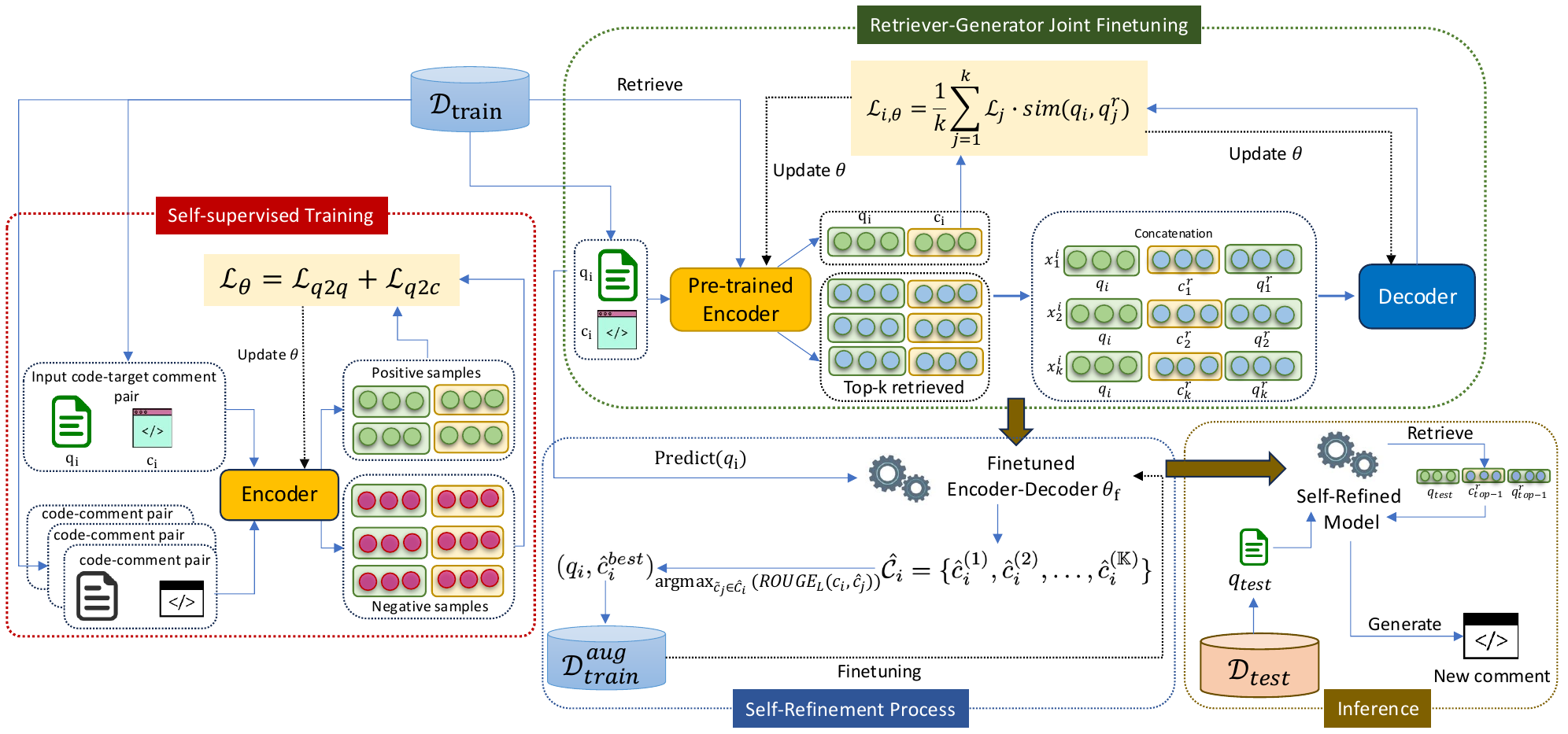}
    \caption{The overall architecture of \tool.}
    \label{fig:architecture}
\end{figure*}

In this paper, we introduce \tool--our proposed approach for Automated Code Comment Generation using Retrieval Augmented Generation, which can enhance the traditional RAG. The overall architecture of \tool is shown in Figure~\ref{fig:architecture}. \tool employs a Encoder-Decoder CodeT5~\cite{wang2021codet5} model with joint fine-tuning to concurrently leverage the performance of Retriever and Generator for code comment generation. 
Our proposed approach consists of three key components: \emph{(i)} Self-Supervised Training of Retriever; \emph{(ii)} Retriever-Generator Joint Fine-tuning; and \emph{(iii)} Self-Refinement Process.

\subsection{Self-Supervised Training of Retriever}
\label{sec:phase1}
Recent research in code retrieval has underscored the value of self-supervised contrastive learning for effective code representation~\cite{parvez2021retrieval,shi2023cocosoda}.
Following this paradigm, we employ a contrastive learning approach to pre-train the encoder of the backbone CodeT5 that captures representations of both code snippets and comments. In particular, we introduce a multi-modal contrastive learning approach to jointly learn representations across the two modalities.
Given a code query $q_i$ and its corresponding comment $c_i$, the CodeT5 encoder first produces two representation vectors, which, for simplicity, are also denoted as $q_i$ and $c_i$, respectively. We fine-tune the encoder to simultaneously enhance both code-to-code and code-to-comment retrieval performance, using {\it in-batch negatives} technique~\cite{liu2020retrieval}. As such, for each training instance $(q_i, c_i)$ in a training batch $\mathcal{B}$, two contrastive loss functions will be computed as follows.
    \begin{equation}
        \mathcal{L}_{q2q} = -\log \frac{e^{\text{sim}(q_i, {q^+_i})/\tau}}{e^{\text{sim}(q_i, {q^+_i})/\tau} + \sum\limits_{\mathcal{B}} e^{\text{sim}({q_i}, {q^-_i})/\tau}}
        \label{eq:loss_code2code}
    \end{equation}
    \begin{equation}
        \mathcal{L}_{q2c} = -\log \frac{e^{\text{sim}({q_i}, {c_i})/\tau}}{e^{\text{sim}({q_i}, {c_i})/\tau} + \sum\limits_{j \in \mathcal{B},j \neq i} e^{\text{sim}({q_i}, {c_j})/\tau}}
        \label{eq:loss_code2comment}
    \end{equation}
In Equation~\ref{eq:loss_code2code}, ${q^+_i}$ denotes the representation vector of the positive code query of $q_i$. In the self-supervised learning setting, the input code query $q_i$ is passed through the encoder twice to produce two representation vectors, ${q_i}$ and ${q^+_i}$. The model is trained to minimize the distance between these two representations while maximizing the distance to other code queries in the batch $\mathcal{B}$, which serve as negative samples (${q^-_i}$). $\text{sim}(\cdot)$ denotes the cosine similarity of two vectors. $\tau$ is the temperature from contrastive learning~\cite{parvez2021retrieval}.
The second loss function, $\mathcal{L}_{q2c}$, aims to minimize the similarity between the input code $q_i$ and its corresponding comment $c_i$, while maximizing the similarity gap with comments from negative samples in the batch.
By pulling semantically aligned code–comment pairs closer and pushing apart misaligned ones, this objective helps the model more effectively distinguish between relevant and irrelevant pairs, thereby enhancing its understanding of the semantic relationship between code and comments.
This self-supervised training phase aims to enhance the encoder's performance on the retrieval task and strengthens the ability of the proposed approach to capture semantic representations, enabling it to generate comments that more accurately reflect the underlying logic of the code.

\subsection{Retriever-Generator Joint Finetuning}
\label{sec:phase2}
We employ the Encoder-Decoder architecture of the CodeT5 backbone to generate summaries for a given code query.
The pre-trained encoder from the previous phase is used to embed code snippets and comments from the training dataset.
Subsequently, we jointly fine-tune both this encoder (for Retriever) and the decoder (for Generator) in a unified training process.
For each code--comment pair $(q_i, c_i)$ in the training set $\mathcal{T}$, the Retriever selects the top-$k$ most relevant code--comment pairs by ranking the cosine similarity between $q_i$ and all other code snippets $q_j \in \mathcal{T}, j \neq i$, resulting in a retrieval set  $\mathcal{R}_i = \{(q^r_1, c^r_1),\ldots,(q^r_k,c^r_k)\}$.
The input query $q_i$ is then concatenated with each retrieval exemplar to form $k$ augmented input sequence: $x^i_j = q_i \oplus  c^r_j \oplus q^r_j$.
$\{x^i_j\}_{j=1}^k$ are then fed into the Decoder to estimate $p(c_i|x^i_j)$ using the Cross Entropy loss function: 
\[\mathcal{L}_j = -\sum\limits_{t=0}\limits^{|c_i|}\log(p(c_i^t|c_{i,<t},x^i_j))\]
In the joint finetuning process, the Retriever will be updated based on the feedback from Generator by taking into account the contribution of each retrieved result. 
The joint loss function $\mathcal{L}_i$ of the input code query $q_i$ is then computed as in Equation~\ref{eq:loss_composite}.
\begin{equation}
    \mathcal{L}_i = \frac{1}{k}\sum\limits_{j=1}\limits^k \mathcal{L}_j \cdot \nu_j
    \label{eq:loss_composite}
\end{equation}
where $\nu_j = \text{sim}(q_i, q^r_j)$ represents the contribution of the retrieved exemplar $q^r_j$ to the generation of the comment $c_i$ of $q_i$.




\subsection{Self-Refinement Process}
\label{sec:phase3}
Auto-regressive sequence generation models commonly suffer from exposure bias problem and hallucination between training and inference phase\cite{to2023better}. To tackle this, we introduce a lightweight post-generation refinement module that improves the faithfulness and fluency of generated comments. For each training example, given an input $q_i$ and its corresponding ground-truth comment $c_i$, we generate $\mathbb{K}$ candidate comments using the joint fine-tuned model $\mathbb{M}$ resulting from the second phase. 
\[\hat{\mathcal{C}}_i = \{ \hat{c}_i^{(1)},\hat{c}_i^{(2)},\ldots,\hat{c}_i^{(\mathbb{K})}\} \text{ where } \hat{c}_i^{(j)} = \mathcal{M}(q_i)\]
We compute the ROUGE--L score between each candidate comment and the reference comment $c_i$. 
The candidate with the highest score is selected to build the augmented dataset $\mathcal{D}_{\text{aug}} = \{(q_i, \hat{c}_i^{best})\}$,which is then used to further fine-tune the joint retrieval–generation model. This allows the model to leverage self-generated, high-quality comments that are most semantically aligned with the ground truth.

\subsection{Inference}
During inference phase, an input code query $q_t$ is only concatenated with the highest retrieval score exemplar $q^t_r, c^t_r$ to generate the comment. 
\[
c_t = \mathcal{M}_{refined}(q_t\oplus c^r_t\oplus q^r_t)
\]

%% file: src/Empirical_Evaluation.tex
We evaluate \tool through a series of experiments on established code summarization benchmarks. 

\subsection{Research Questions}
\label{sec:rq}
\vspace{.2cm}

\noindent \rqone~This research question evaluates the retriever effectiveness of \tool in comparison to baseline methods.

\vspace{.2cm}
\noindent \rqtwo~We compare the efficiency of our approach to baselines. For a fair comparison and consistency in evaluation, we reproduce \JC and \CMR with the same experimental setting as provided in the original studies 
\cite{lu2024improving,li2024cross}.

\begin{itemize}
    \item \CMR~\cite{li2024cross} introduced a joint retriever-generator framework for code summarization, where the retriever and generator are finetuned independently. An extractor is then used to align the retrieved code with the generated comment, refining the final output.
    \item \JC~\cite{lu2024improving} also employed a joint retriever-generator paradigm for comment generation, but treated the retriever and generator as separate models, sharing weights between them during training.
    \item \Llama~\cite{grattafiori2024llama} is a Large Language Model (LLM) developed by Meta AI. Due to resource constraints, we use the 8B-parameter version for inference. In our experiments, the LLM serves as the generator in the RAG framework, with one-shot and few-shot exemplars retrieved using CodeT5 embeddings.
\end{itemize}

\vspace{.2cm}
\noindent \rqthree~We propose strategies to increase model performance, including training of encoder, retriever-generator integration, and a self-refinement mechanism. 
This RQ ascertains 
how each individual component contributes to the overall performance.

\subsection{Benchmark Datasets}
\label{sec:dataset}
We evaluate our approach on \JCSD, \PCSD and \CCSD--the most popular benchmark datasets for code summarization. Specifically, the Java dataset \cite{hu2018deep} comprises pairs of source code and corresponding comments from well-known GitHub repositories, the Python dataset initially gathered by Baron et al. \cite{barone2017parallel}, the dataset \JCSD and \PCSD was preprocessed by Lu et al. \cite{lu2024improving} to remove duplication. The C Code  dataset (\CCSD) was crawled by Liu et al. \cite{liu2020retrieval} with 95k function-summary pairs. The statistics of datasets are shown in Table \ref{tab: datasets}.
\renewcommand{\arraystretch}{1.2}
\begin{table}[t!]
    \centering
    \setlength{\tabcolsep}{12pt}
    \caption{Statistic of the datasets.}
        \begin{tabular}{l|c|c|c}
             \hline 
             Dataset& JCSD & PCSD &CCSD\\
             \hline
             Training set & 69,708 & 55,538 & 84,315\\
             Validation set & 8,714 & 18,505 & 4,432\\
             Testing set & 6,489 & 18,142 & 4,203\\ \hline
        \end{tabular}
    \label{tab: datasets}
\end{table}

\subsection{Evaluation metrics}
Following prior work \cite{li2024cross, lu2024improving, gao2022m2ts}, we evaluate \tool using BLEU \cite{papineni2002bleu}, ROUGE-L \cite{lin2004rouge}, METEOR \cite{banerjee2005meteor}, and CIDER \cite{vedantam2015cider}. Corpus-level BLEU captures overall performance while Sentence-level BLEU evaluates individual predictions. ROUGE-L evaluates the similarity between generated and reference texts using the longest common subsequence. METEOR offers improvements over traditional metrics by considering linguistic aspects such as synonymy, stemming, and word order. CIDEr computes the relevance of key information.\footnote{Due to space limitations, we omit the details of these metrics.}

\renewcommand{\arraystretch}{1.4}
\begin{table*}[t]
\centering
\caption{Comparison of \tool with the baselines.}
\resizebox{\textwidth}{!}{%
\begin{tabular}{l|ccccc|ccccc|ccccc}
\hline
\multirow{2}{*}{\textbf{Approach}} 
& \multicolumn{5}{c|}{JCSD} 
& \multicolumn{5}{c|}{PCSD} 
& \multicolumn{5}{c}{CCSD} \\
\cline{2-16}
& C-B & S-B & RL & M & C & C-B & S-B & RL & M & C & C-B & S-B & RL & M & C \\

\hline
Llama3.1\textsubscript{RAG 1-shot} \cite{grattafiori2024llama}  & 15.08 & 14.61 & 35.26 & 18.33 & 1.69 & 11.66 & 7.08 & 21.28 & 14.09 & 0.96 & 16.46 & 12.09 & 34.01 & 19.27 & 1.76 \\
Llama3.1\textsubscript{RAG n-shot} \cite{grattafiori2024llama}& 15.15 & 14.51 & 36.3 & 18.97 & 1.74 & 20.89 & 13.28 & 35.96 & 24.13 & 1.85 & 19.65 & 13.38 & 37.57 & 20.55 & 2.04 \\
CMR-Sum\cite{li2024cross} & 23.53 & 23.24 & 46.59 & 20.5 & 2.7 & 28.89 & 22.41 & 52.42 & 23.94 & 2.9& 21.72 & 15.85 & 40.96 & 19.8 & 2.45 \\
JOINTCOM\cite{lu2024improving} & 26.09 & 26.53 & 50.22 & 22.02 & 2.99 & 27.89 & 21.05 & 52.6 & 23.84 & 2.82 & 26.32 & 19.99 & 46.15 & 22.82 & 2.91 \\
\textbf{RAGSum} & \textbf{27.16} & \textbf{27.94} & \textbf{51.54} & \textbf{22.71} & \textbf{3.13} & \textbf{33.0} & \textbf{26.11} & \textbf{56.15} & \textbf{26.53} & \textbf{3.28} & \textbf{27.95} & \textbf{21.36} & \textbf{47.35} & \textbf{23.76} & \textbf{3.03}\\
\hline
\end{tabular}%
}
\label{tab:baseline_results}
\end{table*}

\renewcommand{\arraystretch}{1.4}
\begin{table*}[t]
\centering
\caption{Ablation study.}
\resizebox{\textwidth}{!}{%
\begin{tabular}{l|ccccc|ccccc|ccccc}
\hline
\multirow{2}{*}{\textbf{Approach}}  & \multicolumn{5}{c|}
{JCSD} & \multicolumn{5}{c|}{PCSD} & \multicolumn{5}{c}{CCSD} \\
\cline{2-16}
& C-B & S-B & RL & M & C & C-B & S-B & RL & M & C & C-B & S-B & RL & M & C \\

\hline
Only Generator & 13.33 & 14.16 & 41.3 & 15.63 & 1.88 & 21.42 & 15.36 & 48.64 & 21.3 & 2.28 & 18.28 & 12.71 & 39.13 & 18.83 & 2.16 \\
RAGSum \textsubscript{\textit{w/o} combined} & 24.02 & 24.03 & 48.37 & 20.92 & 2.77 & 28.69 & 22.07 & 52.64 & 24.24 & 2.88 & 23.1 & 17.23 & 42.89 & 21.13 & 2.59 \\
RAGSum \textsubscript{\textit{w/o} pretrained + SR} & 27.12 & 27.25 & 50.59 & 22.47 & 3.05 & 31.69 & 24.8 & 55.04 & 25.73 & 3.16 & 27.19 & 20.84 & 46.75 & 23.27 & 2.96 \\
RAGSum \textsubscript{\textit{w/o} SR} &27.06 & 27.8 & 51.24 & 22.62 & 3.1 & 32.57 & 25.66 & 55.65 & 26.22 & 3.23 & 27.76 & 21.26 & 47.18 & 23.54 & 3.02 \\

\textbf{RAGSum} & \textbf{27.16} & \textbf{27.94} & \textbf{51.54} & \textbf{22.71} & \textbf{3.13} & \textbf{33.0} & \textbf{26.11} & \textbf{56.15} & \textbf{26.53} & \textbf{3.28} & \textbf{27.95} & \textbf{21.36} & \textbf{47.35} & \textbf{23.76} & \textbf{3.03}\\
\hline
\end{tabular}%
}
\label{tab:ablation_results}
\end{table*}

\subsection{Implementation Details}
\tool is built on the pre-trained CodeT5-base model. We use a batch size of 24, learning rates of $5 \times 10^{-5}$ for fine-tuning and $1 \times 10^{-5}$ for self-improvement, and temperature $\tau = 0.2$. Training includes 1 pretraining epoch, 10 fine-tuning epochs, and 5 self-refinement epochs.

%% file: src/Result.tex
\subsection{\rqone}
\begin{figure}[H]
    \centering
    \includegraphics[width=\linewidth]{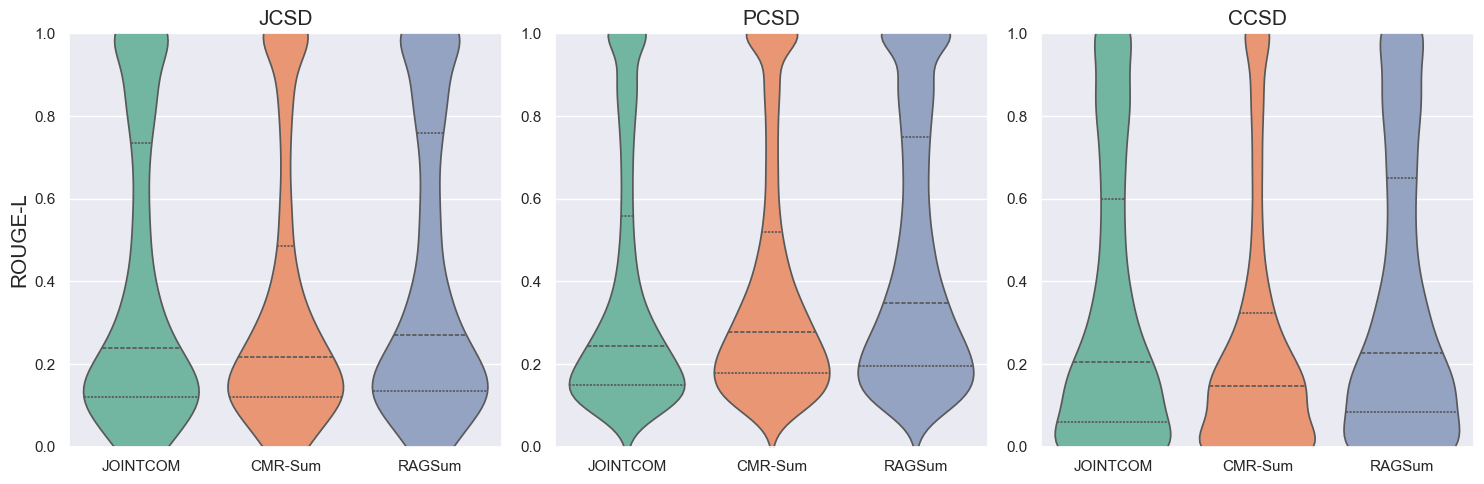}
    \caption{Distribution of Retrieved Comments and Targets Across Methods.}
    \label{fig:violin_retriever}
\end{figure}

We used the retrievers of \tool, \JC, and \CMR to fetch relevant code and calculate ROUGE-L scores against ground truth comments. Figure~\ref{fig:violin_retriever} shows that \tool consistently achieves higher median scores across all datasets. On \PCSD, \tool achieves the highest median at 0.37, followed by \CMR at 0.3 and \JC at 0.25.
The upper quartile of \tool extends beyond 0.75, while \JC and \CMR are remain below this level on \JCSD dataset. Overall, \tool’s score distribution shifts upward, indicating more relevant top-1 retrieved comments. This highlights the effectiveness of the joint fine-tuning process in enhancing the capability of our retriever.
\vspace{.2cm}
\noindent\fbox{
\begin{minipage}{0.98\columnwidth}
		\textbf{Answer to RQ$_1$:}
The retriever 
is more effective and robust than the baseline methods in fetching relevant information.
\end{minipage}}

\subsection{\rqtwo}
We compare our approach to the baselines on three datasets, with results summarized in Table \ref{tab:baseline_results}. The metrics include \textbf{C-B} (Corpus-BLEU), \textbf{S-B} (Sentence-BLEU), \textbf{RL} (ROUGE-L), \textbf{M} (METEOR), and \textbf{C} (CIDEr). In Java dataset, compared to the best baselines \JC, \tool increases 4.1\%, 5.31\%, 2.63\%, 3.13\% and 4.68\% in terms of C-BLEU, S-BLEU, ROUGE-L, METEOR, and CIDEr, respectively. With \PCSD, our approach significantly outperforms, \tool achieves 33.0, 26.11, 56.15, 26.53 and 3.28 points with improvements of 14.23\%, 16.51\%, 7.12\%, 10.82\%, and 13.1\%, respectively, compared to \CMR. These gains reflect the enhanced alignment between the retriever and generator, which enables more accurate and semantically relevant summary generation For the C dataset, the performance of \tool remains competitive. While the margins over \JC are narrower due to the inherently lower redundancy and more complex structure of C programs, \tool still achieves a gain of 
2.6\% in ROUGE-L and a 6.19\% boost in C-BLEU, suggesting its strong generalization even under challenging conditions.

Moreover, across three benchmarks, \tool consistently outperforms \Llama in both 1-shot and n-shot configurations. Notably, \tool leverages relevant knowledge to generate more context-aware comments.
Overall, the results in Table \ref{tab:baseline_results} demonstrate the superior performance and strong generalization capabilities of \tool across diverse programming languages. The combination of joint fine-tuning, encoder pretraining, and self-improvement enables \tool to effectively model the structural and semantic complexity of code, setting a new state-of-the-art in comment generation.

\vspace{.2cm}
\noindent\fbox{
\begin{minipage}{0.98\columnwidth}
		\textbf{Answer to RQ$_2$:}
On the three given datasets, \tool substantially outperforms the considered baselines with respect to all the evaluation metrics.   
\end{minipage}}

\subsection{\rqthree}
We conduct an ablation study to assess the contribution of key components through four settings: (1) \textbf{Only Generator} fine-tuning CodeT5 without relevant code-comment pairs; (2) \textbf{RAGSum\textsubscript{\textit{w/o} combined}} using retriever and generator independently; (3) \textbf{RAGSum\textsubscript{\textit{w/o} pretrained + SR}} removing both pre-trained encoder and self-refinement process; (4) \textbf{RAGSum\textsubscript{\textit{w/o} SR}} excluding self-refinement. Results are shown in Table \ref{tab:ablation_results}.
Notably, RAGSum\textsubscript{\textit{w/o} combined} exhibits a significant decrease of 13.07\% for Java, 15.02\% for Python, and 20.99\% for C in the Corpus-BLEU metric, primarily due to the absence of the joint fine-tuning strategy, which is essential for effectively aligning the retriever and generator components. It can be observed that excluding both encoder pretraining and the self-improvement mechanism consistently degrades the model’s performance across all metrics and programming languages. Further analysis shows that removing the self-refinement mechanism results in a performance degradation across all metrics. For instance, in \PCSD, Corpus-BLEU falls from 33.0 to 32.57, and in \CCSD, it drops from 27.95 to 27.76.

\begin{figure}[t]
    \centering
    \includegraphics[width=0.5\textwidth]{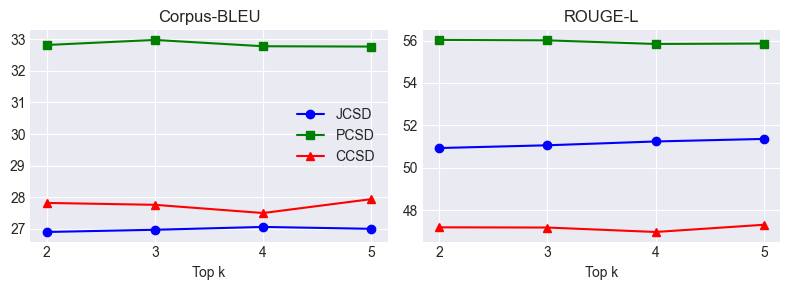}
    \caption{Top-$k$ Impact Scores.}
    \label{fig:top-k}
\end{figure}

For a more in-depth analysis of the effectiveness of top $k$ relevant code comment pairs during fine-tuning, Figure~\ref{fig:top-k} compares top-$k$ values using C-BLEU and ROUGE-L. Optimal performance is at $k$ value 4 for \JCSD and 3 for both \PCSD and \CCSD.

\vspace{.2cm}
\noindent\fbox{
\begin{minipage}{0.98\columnwidth}
		\textbf{Answer to RQ$_3$:}
Each component of \tool contributes significantly to its overall performance, enhancing different aspects of the model.
\end{minipage}}

\subsection{Qualitative analysis}
\begin{figure}[t!]
    \centering
    \includegraphics[width=0.96\linewidth]{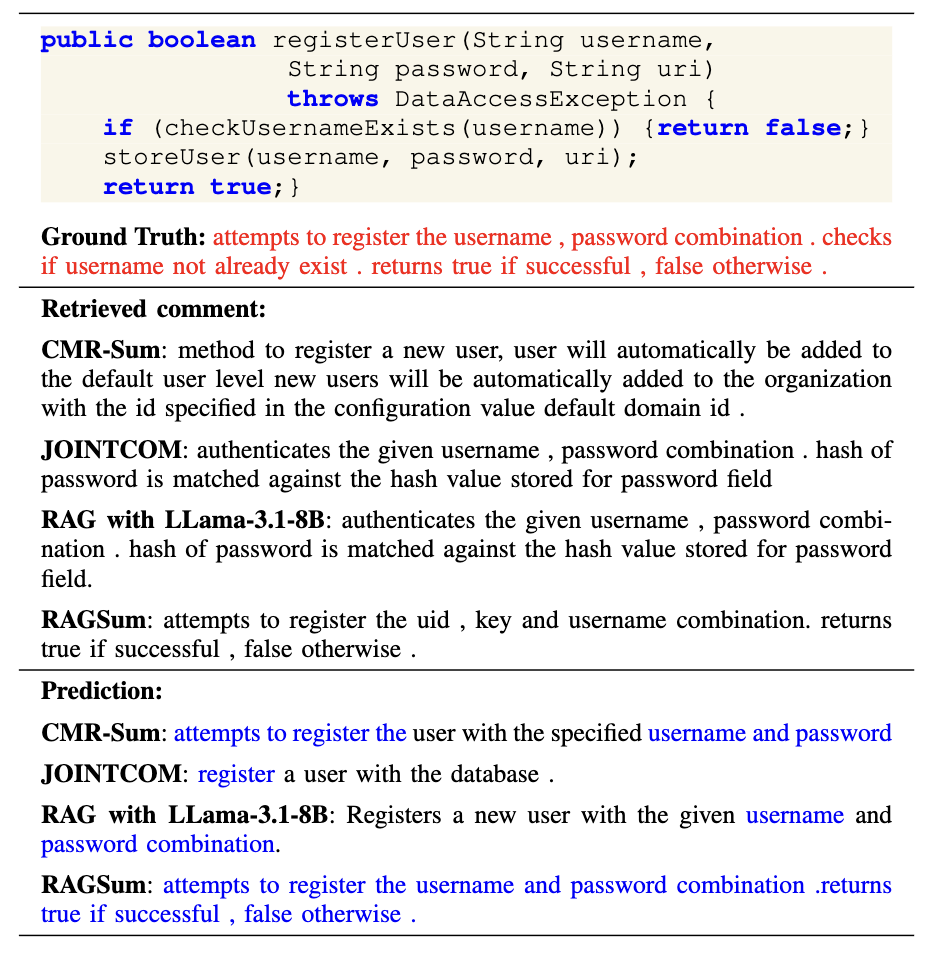}
    \caption{Example of Java code.}
    \label{fig:java_example}
\end{figure}

\begin{figure}[t!]
    \centering
    \includegraphics[width=0.9\linewidth]{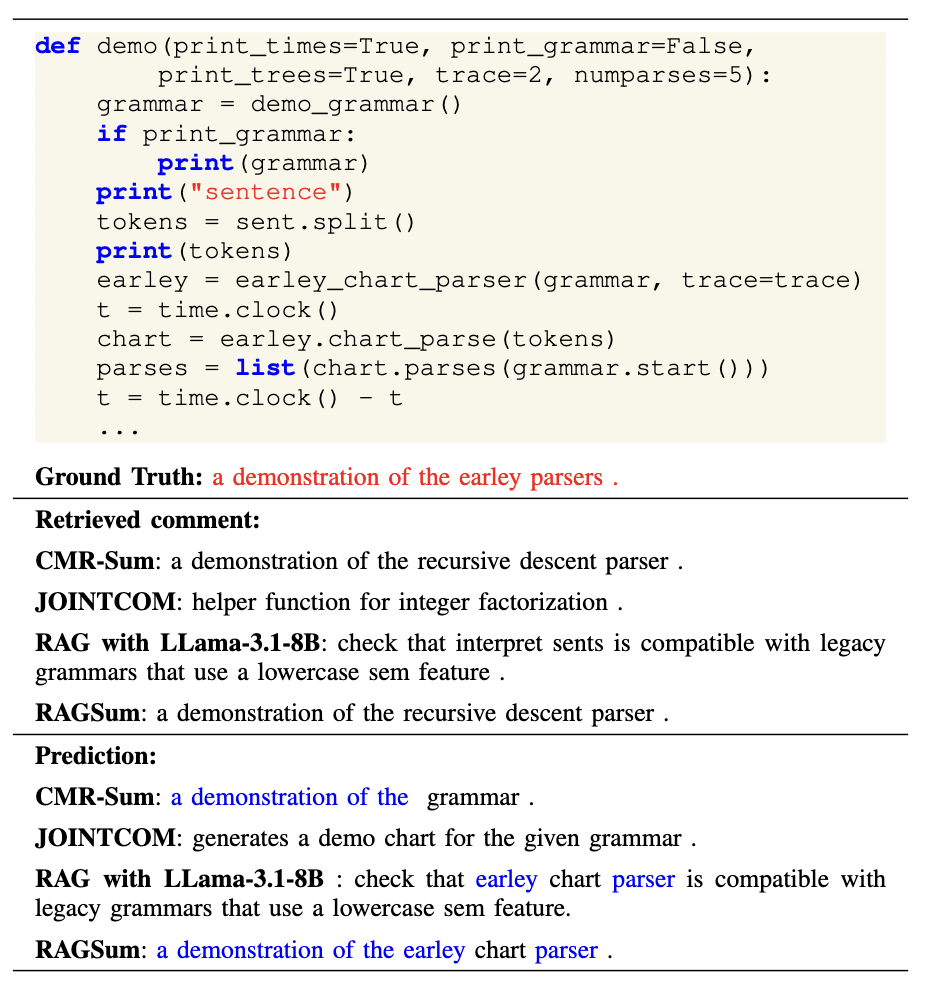}
    \caption{Example of Python code.}
    \label{fig:python_example}
\end{figure}

We present two examples demonstrating the superior efficiency of our retrieval method compared to existing baselines. This highlights the impact of relevant code comment on the quality of generated comment. Figure \ref{fig:java_example} shows the example in Java code, \tool retrieves comment which is the most relevant to target, provide meaningful context for generation. In this case, the comments are generated by \CMR and \JC lack the information contained in the code snippet. In Fig \ref{fig:python_example}, \CMR retrieves the same comment with \tool, but the generated comment by \CMR fails to align closely with target, because our approach employs a joint modeling mechanism that better integrated retrieval and generation. \JC 's retriever performance is limited in Fig \ref{fig:python_example}, leading to a generated comment that lacks sufficient information from the given code. 
In both cases, \Llama uses relevant code and comments but still produces outputs misaligned with the reference. 

\subsection{Threats to Validity}

\smallskip
\noindent
$\triangleright$ \textbf{Internal validity.}
We used the most popular metrics for evaluating
code summarization but it may have some limitations. These metrics may not fully capture semantic equivalence, potentially underestimating the quality. Therefore, it is necessary to evaluate generated summaries from additional perspectives, such as
human evaluation \cite{hu2020deep}.

\smallskip
\noindent
$\triangleright$ \textbf{External validity.} Potential threat to validity lies in the variation of results and performance of our approach with different coding styles, programming languages and levels of complexity. To mitigate this, we selected three widely used datasets with different programming languages, aiming to capture a broad range of code characteristics.

%% file: src/Conclusion.tex
In this paper, we proposed \tool for automated code comment generation that effectively leverages the existing joint fine-tuning retriever and generator. Empirical evaluation of benchmark datasets showed that \tool significantly improved baselines in code summarization. For future work, we plan to explore more dynamic retrieval mechanisms, investigate the scalability of \tool to large-scale codebases, and extend our approach to support multilingual codebases and more diverse programming paradigms. 